\documentclass[
trackchanges,
longbib
]{aastex701}

\usepackage{graphicx} 
\usepackage{adjustbox, amsmath}
\usepackage{aas_macros}
\usepackage{xfrac}

\begin{document}

\title{Electrical coupling of a horizontal dipole antenna to a dielectric half-space: applications to radio astronomy from the lunar surface}

\author[0000-0002-9181-9948]{Kaja M. Rotermund}
\affiliation{Physics Division, Lawrence Berkeley National Laboratory, Berkeley, CA 94720, USA}
\email{krotermund@lbl.gov}

\author[0000-0001-8101-468X]{Aritoki Suzuki}
\affiliation{Physics Division, Lawrence Berkeley National Laboratory, Berkeley, CA 94720, USA}
\email{asuzuki@lbl.gov}


\author[0000-0002-1989-3596]{Stuart D. Bale}
\affiliation{Physics Department, University of California, Berkeley, CA 94720-7300, USA}
\affiliation{Space Sciences Laboratory, University of California, Berkeley, CA 94720-7450, USA}
\affiliation{The Blackett Laboratory, Imperial College London, London, SW7 2AZ, UK}
\email{bale@berkeley.edu}

\author[0000-0002-8713-3695]{An\v{z}e Slosar}
\affiliation{Brookhaven National Laboratory, Upton, NY 11973, USA}
\email{anze@bnl.gov}

\date{February 2026}

\newcommand{\Tsky}{T_\mathrm{sky}}
\newcommand{\TA}{T_\mathrm{A}}
\newcommand{\Tmoon}{T_\mathrm{moon}}
\newcommand{\fground}{F_\mathrm{ground}}
\newcommand{\fsky}{F_\mathrm{sky}}

\newcommand{\R}{R_\mathrm{r}}
\newcommand{\X}{X_{\mathrm{A}}}
\newcommand{\Zant}{Z_\mathrm{A}}
\newcommand{\Zrec}{Z_\mathrm{rec}}
\newcommand{\GammaVD}{\Gamma_\mathrm{VD}}
\newcommand{\Voc}{V_{\mathrm{oc}}}
\newcommand{\Vout}{V_{\mathrm{out}}}
\newcommand{\CB}{C_{\mathrm{B}}}
\newcommand{\Rout}{R_{\mathrm{out}}}

\begin{abstract}
The far side of the Moon, shielded from terrestrial radio frequency interference and beyond the influence of Earth's ionosphere, should offer a uniquely quiet environment for radio astronomy and cosmological experiments. The radio sky below 30 MHz is largely unexplored and is thought to contain spectral signatures of new physics in the early, high-redshift Universe. Achieving precision measurements in this band requires accurate understanding of antenna performance and systematics. For upcoming lunar surface radio astronomy missions, this modeling will be challenging because antennas will deploy at heights that are only a small fraction of a wavelength above the lunar regolith, where strong coupling between the antenna and the surface can significantly alter impedance, radiation patterns, and efficiency. The challenge is compounded by the layered dielectric structure of the regolith and the tendency for permittivity to increase with depth, both of which are difficult to represent faithfully in numerical simulations.

In this work, we review theoretical predictions for the behavior of a simple horizontal dipole above a dielectric half-space, representing the lunar regolith, and compare them with simulation results obtained using the Ansys HFSS integral equation solver. We quantify how the antenna impedance and beam pattern couple to the sky for a representative lunar surface radio astronomy experiment. The results show that surface induced effects decrease rapidly, even for modest increases in antenna height above the regolith. Conversely, a dipole antenna placed on or very near the lunar surface will exhibit complex spectral response that renders systematics control very difficult without detailed information on regolith properties.
\end{abstract}

\section{Introduction}
The lunar far-side is a uniquely radio-quiet environment from which exquisitely sensitive measurements can be conducted at low-radio frequencies; to-date a largely unexplored regime. Ground-base experiments suffer from two key limitations: the Earth's ionosphere which is opaque to radiation below 30~MHz and terrestrial and man-made radio frequency interference (RFI). Access to the lunar far-side is becoming more readily available with the advent of the commercial lunar payload services (CLPS), a NASA initiative partnering with American companies to bring science payloads to the lunar surface.  The LuSEE-Night instrument \citep{BaleLSN} targeting the 1~MHz -- 50~MHz radio-frequency sky will be deployed on the lunar far-side by CLPS in 2027.  

While measurements conducted by the Radio Astronomy Explorer 2 Satellite \citep{RAE1975} in the 70's identified the lunar far-side as a pristine radio-quiet environment, deploying to the lunar surface imposes several challenges. One critical consequence of landing on the moon is the effect of the regolith on the antenna response. An antenna some finite distance from a dielectric ground will capacitively couple to the ground, diminishing the skyward facing gain. The degree to which the antenna response is affected by the ground beneath it depends on several variables, including but not limited to the height of the antenna above the ground and the dielectric properties of the ground. These are two parameters we consider in this work.

The dielectric properties of the lunar regolith are still largely unknown, especially on the far-side. A re-analysis of the Apollo 17 Mission provides some insight into how the regolith changes with depth \citep{Grimm2018}. However, these measurements were conducted in the Taurus-Littrow valley of the lunar near-side. Due to the tidally-locked nature of the Moon, it is entirely appropriate to assume the far-side regolith will differ from the near-side. Despite this, the Apollo 17 measurements and \citet{Grimm2018} analysis provide insight into reasonable values to consider. In this work we consider a simple dipole antenna 6-meters tip-to-tip some height above an infinite dielectric half-space to better understand the effects on the dipole. We compare analytical model to a EM simulations using the Ansys High Frequency Simulation Software (HFSS) integral equation (IE) solver \citep{HFSS}. The IE solver allows for the study of the beam both towards the sky as well as into the ground. We then quantify how the antenna impedance and beam couple to the sky for a representative lunar surface radio astronomy experiment.

\section{Analytical Model}
The radiated field of an electrical dipole over a dielectric half-space has long been derived and is further expanded to a general multi-layer ground structure \citep{SEP, Cooper, Wait}. The general approach to solving this problem consists of treating the Transverse Electric (TE) and Transverse Magnetic (TM) radiated fields from a horizontal electric dipole antenna separately and obtaining an integral form for the radiated fields. Then, asymptotic expansion of the fields are calculated to obtain their far-field radiation patterns \citep{SEP}. 

We specifically followed the derivation by \citet{Jiao}, as they give a closed form for the TE and TM radiated fields of a half-wave horizontal dipole antenna at a height $h$ above a lossless dielectric half-space. 
Assuming the dipole lies along the $x$-axis, the TE-field approximation is given by:
\begin{equation}
    \begin{aligned}
         \left|E_{\phi}^{z>0}(\theta, \phi)\right|  \sim & \left| 1 + \frac{\cos(\theta)-\sqrt{\epsilon_r-\sin^2(\theta)}}{\cos(\theta) + \sqrt{\epsilon_r - \sin^2(\theta)}}e^{i4\pi h \cos(\theta)}\right|  \sin(\phi) \\
        \left|E_{\phi}^{z<0}(\theta, \phi)\right|  \sim 2\epsilon_r^{0.25} &\left|\frac{\cos(\theta)}{-\sqrt{\epsilon_r}\cos(\theta) + \sqrt{\epsilon_r - \sin^2(\theta)}}e^{i2\pi h \sqrt{1-\epsilon_r\sin^2(\theta)}}\right|\sin(\phi),
    \end{aligned}
    \label{eq:theoryMath}
\end{equation}
where we follow polar-coordinate convention: angle $\theta$ is measured down from the $+z$ axis and angle $\phi$ is measured from the $+x$ axis in the $xy-$plane. $\epsilon_r$ is the relative dielectric constant of the half-space. 
As noted by \citet{Jiao}, this formalism is an approximation and should only be used to inform qualitative trends. This is especially true for the model of the TM field, which has larger discrepancies with measurements than the TE field. The TM field formalism is analogous to the TE field and can be referenced in \citet{Jiao}.

To determine the power radiated 
we use the Poynting theorem in the far-field regime~\citep{SEP}. Specifically, the power radiated into the ground and into the sky, respectively, is given by:
\begin{equation}
    \begin{aligned}
    P_{sky} \propto Z_0\left(\int_0^{2\pi}\int_0^{\pi/2}|E_{\phi}^{z>0}(\theta,\phi)|^2\sin(\theta)d\theta d\phi + \int_0^{2\pi}\int_0^{\pi/2}|H_{\phi}^{z>0}(\theta,\phi)|^2\sin(\theta)d\theta d\phi\right) \\
    P_{ground} \propto Z\left(\frac{\epsilon_r}{\mu_r}\int_0^{2\pi}\int_0^{\pi/2}|E_{\phi}^{z<0}(\theta,\phi)|^2\sin(\theta)d\theta d\phi + \int_0^{2\pi}\int_0^{\pi/2}|H_{\phi}^{z<0}(\theta,\phi)|^2\sin(\theta)d\theta d\phi\right).
    \end{aligned}
    \label{eq:PowerRad}
\end{equation}
Here, $Z=\sqrt{\sfrac{\mu}{\epsilon}}$ is the impedance of a  medium and $Z_0$ is the special case for free-space. 



Finally, we define a term, $F_{sky}$, to denote the ratio of the beam into the ground vs the total beam as following: 
\begin{equation}
    F_{sky}=\frac{P_{sky}}{P_{sky} + P_{ground}}.
    \label{eq:fground}
\end{equation}

\section{Simulation}
\begin{figure}[h]
    \centering
    \includegraphics[height=0.35\linewidth]{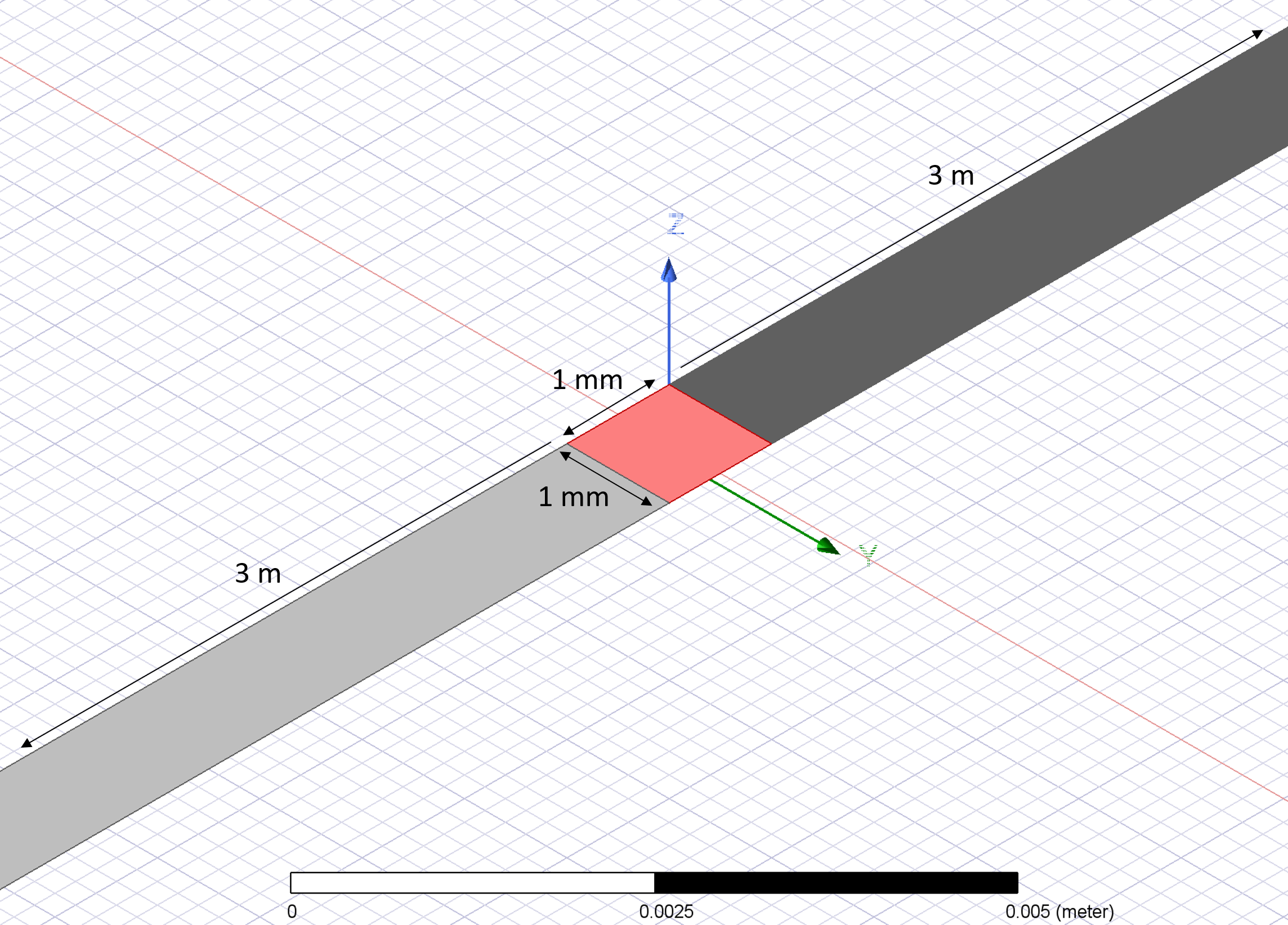}
    \caption{Isometric view of the HFSS simulation. The dipole antenna consists of two 3-meter long perfect electrical conductor strips that are 1~mm wide. The two monopoles are separated by 1~mm and are excited by a 50~$\Omega$ lumped-port, shown in red.}
    \label{fig:simview}
\end{figure}

The integral equations (IE) solver function of Ansys HFSS \citep{HFSS} was used to simulate a center-fed horizontal dipole above an infinite half-space. All relevant dimensions are shown in Fig.~\ref{fig:simview}. The monopoles are perfect electric conductor sheets each 3~meters in length and 1~mm wide. They are separated by a 1~mm 50~$\Omega$ lumped-port excitation. The dipole is oriented along the $x$-axis and is a height $h$ above an infinite dielectric half-space, which is defined to have a dielectric constant of $\epsilon_r =4$ and no loss ($\tan \delta = 0$). The far-field antenna response was simulated between 1~MHz -- 50~MHz at 1~MHz increments. We conducted a parameter sweep, raising the height from 0 to 5 meters. All results were obtained with a maximum mesh size of 10~mm on the dipole surface. 

\begin{figure}
  \centering
  \includegraphics[height=5.3cm]{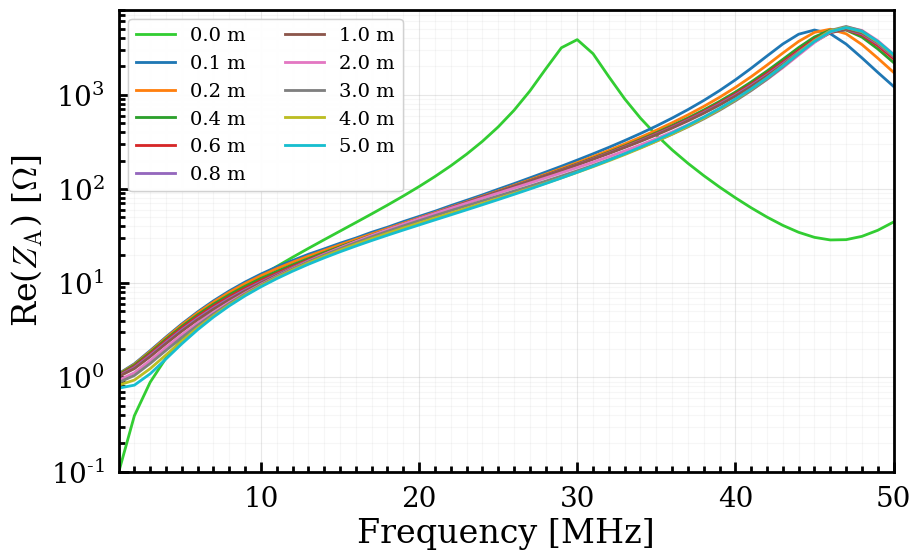}
  \includegraphics[height=5.3cm]{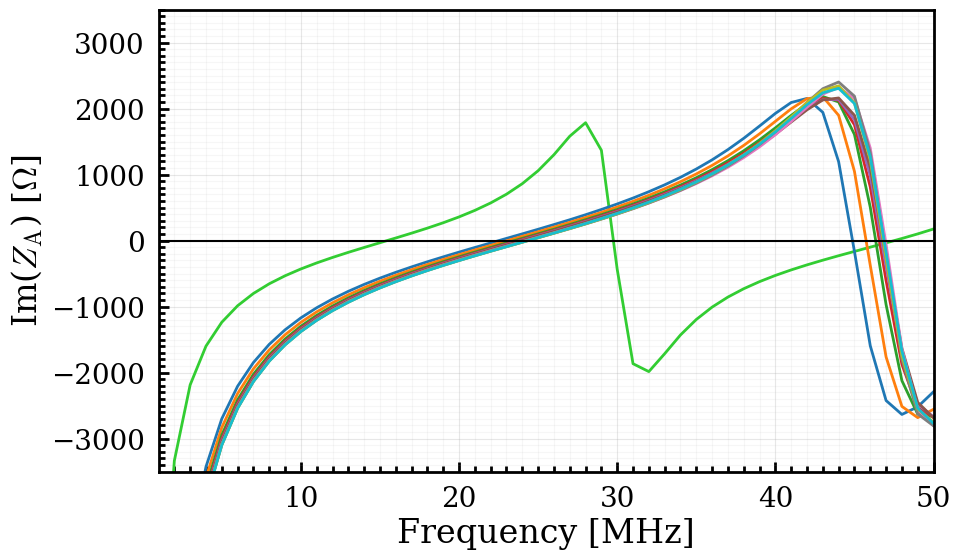}
  \caption{Real (\emph{left}) and imaginary (\emph{right}) components of the complex antenna impedance $Z_A$ as simulated for a simple 6-meter tip-to-tip dipole some height above a dielectric half-space.
  \label{fig:ZA}}
\end{figure}

\begin{figure}
  \centering
  \includegraphics[height=5.3cm]{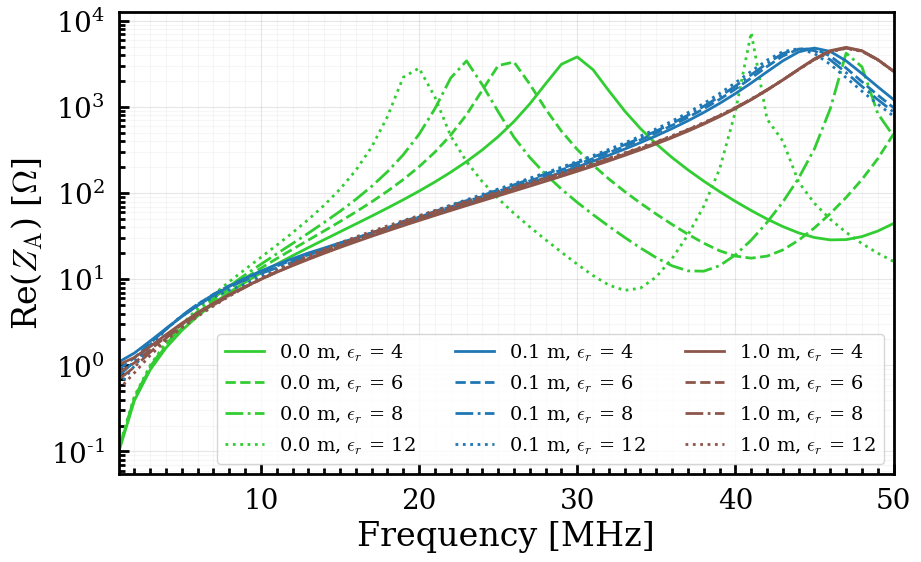}
  \includegraphics[height=5.3cm]{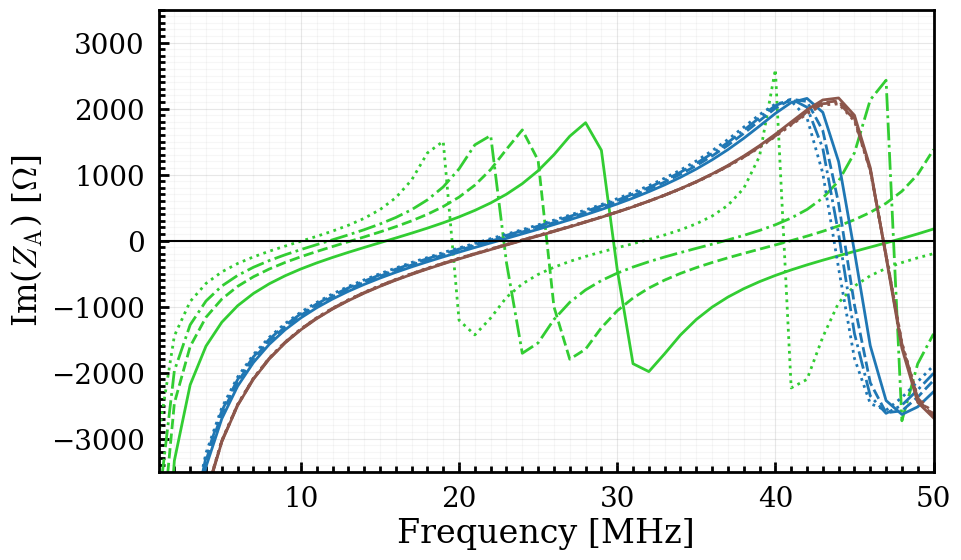}
  \caption{Real (\emph{left}) and imaginary (\emph{right}) components of the complex antenna impedance as simulated for a simple 6-meter tip-to-tip dipole 0~meter (green), 0.1~meter (blue), and 1~meter (brown) high with varying dielectric constants.
  \label{fig:ZA_Er_comp}}
\end{figure}

\section{Results}
The simulated impedance for the dipole antenna is shown in Fig.~\ref{fig:ZA}. Generally, the impedance is largely unaffected by the height. However, when the dipole is close to the dielectric half-space, i.e. less than about 20~cm, the dipole's impedance is modified by the presence of the dielectric half-space. For the unique case when the dipole is at 0-meters, the frequency of the peak real-impedance shifts by over 15~MHz. 

We consider the unique case of the dipole touching the half-space in more detail. We simulated the same dipole with changes to the dielectric constant, increasing it from $\epsilon_r=4$ to 6, 8, and 12. As is seen in Fig.~\ref{fig:ZA_Er_comp} (green lines), the dielectric constant has a strong effect on the impedance when it is physically touching the dielectric half-space. The peak impedances shifts towards lower frequencies as $\epsilon_r$ is increased. As a reference, we simulated the dipole with the same dielectric constants this time 0.1~meters and 1~meter above the half-space. For the 1~meter case (brown lines) we find changes to the dielectric constant to have a negligible effect on the impedance. For the 0.1~meter case (blue lines), we see only small changes to the impedance, suggesting that even small heights above the half-space already suppresses this dependency on dielectric constant substantially.

\begin{figure}
    \centering
    \includegraphics[width=0.8\linewidth]{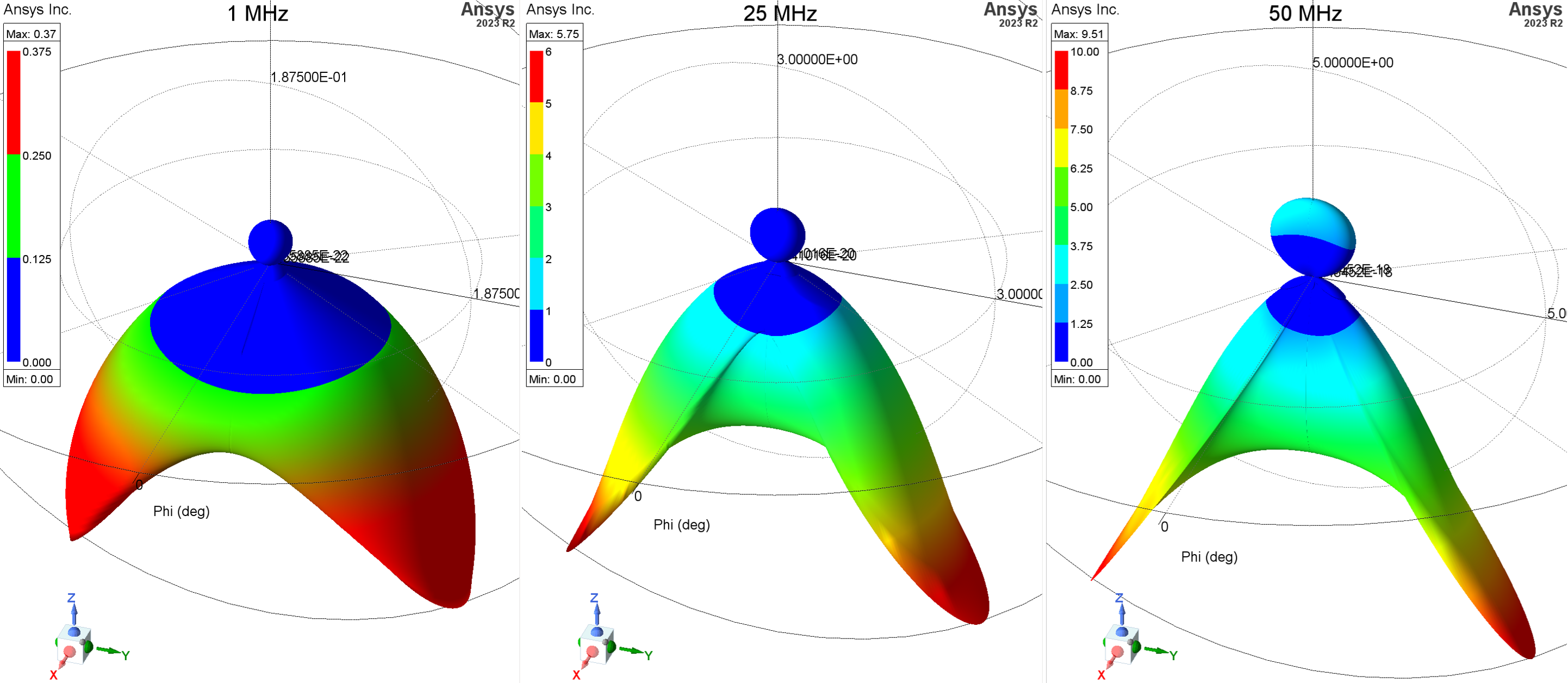}
    \caption{Representative 3D antenna gain plots simulated by HFSS at 1~MHz, 25~MHz, and 50~MHz for a 6-meter tip-to-tip dipole 0.5~meters above a dielectric half-space with dielectric constant $\epsilon_r=4$.}
    \label{fig:3DTEPower}
\end{figure}

\begin{figure}
    \centering
    \includegraphics[width=\linewidth]{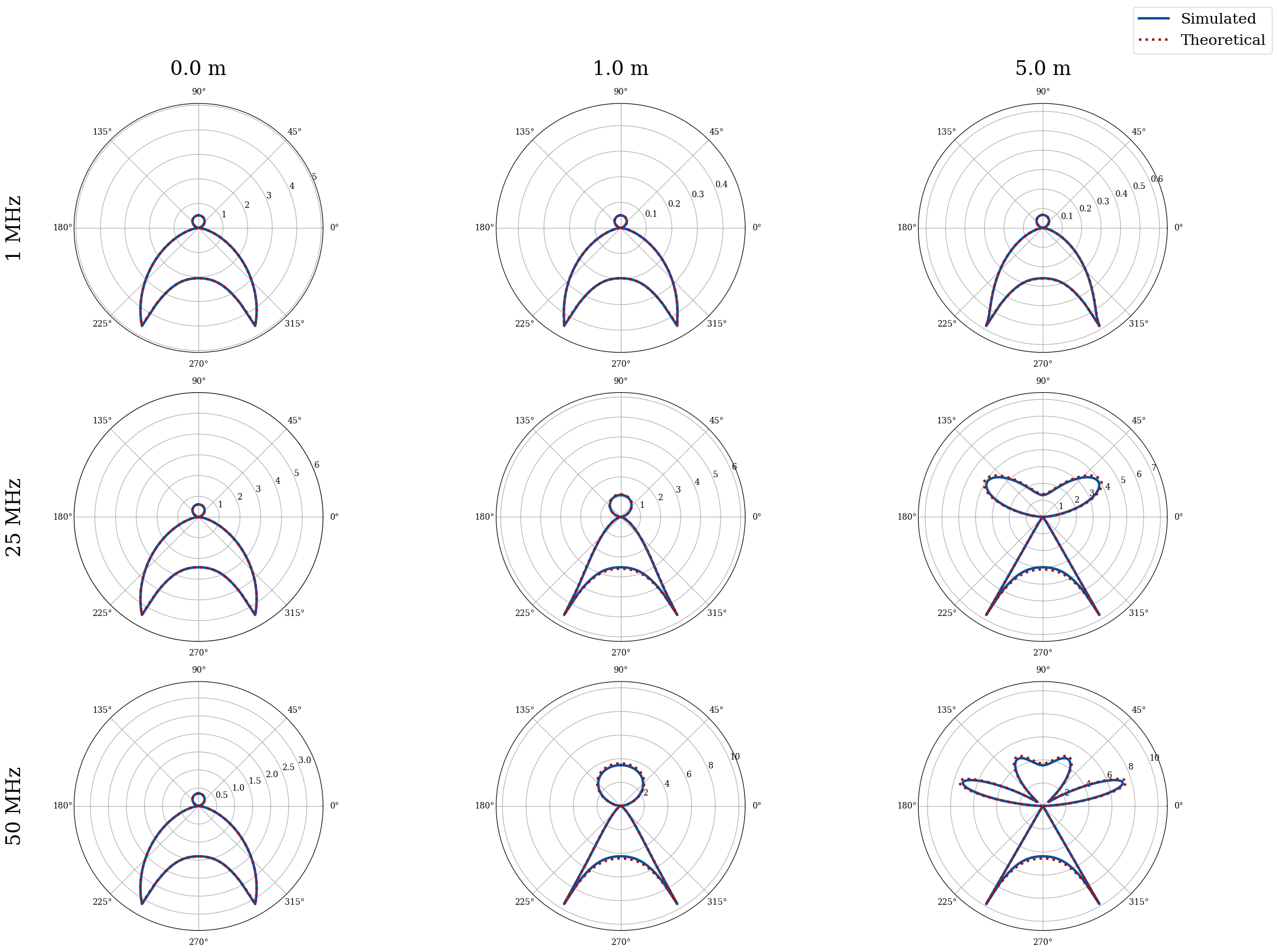}
    \caption{Representative radiation plots at 1~MHz, 25~MHz, and 50~MHz for a 6-meter tip-to-tip dipole 0~meters, 1~meter, and 5~meters above a dielectric half-space with dielectric constant $\epsilon_r=4$. The blue lines are the extracted H-plane simulated by HFSS. The red dashed lines are theoretical predictions (TE) from the formalism defined in text.}
    \label{fig:TEPower}
\end{figure}

Representative simulated 3D antenna beams at 1~MHz, 25~MHz, and 50~MHz at 0.5~meters height are shown in Fig.~\ref{fig:3DTEPower}. In Fig.~\ref{fig:TEPower}, we show representative radiation patterns of the H-plane ($\phi=90^{\circ}$). Both HFSS simulations (blue lines) and theoretical predictions (dotted red curves) are included, showing agreement between the two. Thus, we use simulated result for further analysis below. At low heights, the majority of the beam is in the dielectric ground and the skyward facing beam is a relatively simple sphere. At higher heights, the gain of the skyward facing beam increases, however, at the same time the complexity of the beam shape increases. The trade-off between antenna sensitivity and beam chromaticity is well established.




\section{Analysis}
\begin{figure}
  \centering
  \includegraphics[height=4.2cm]{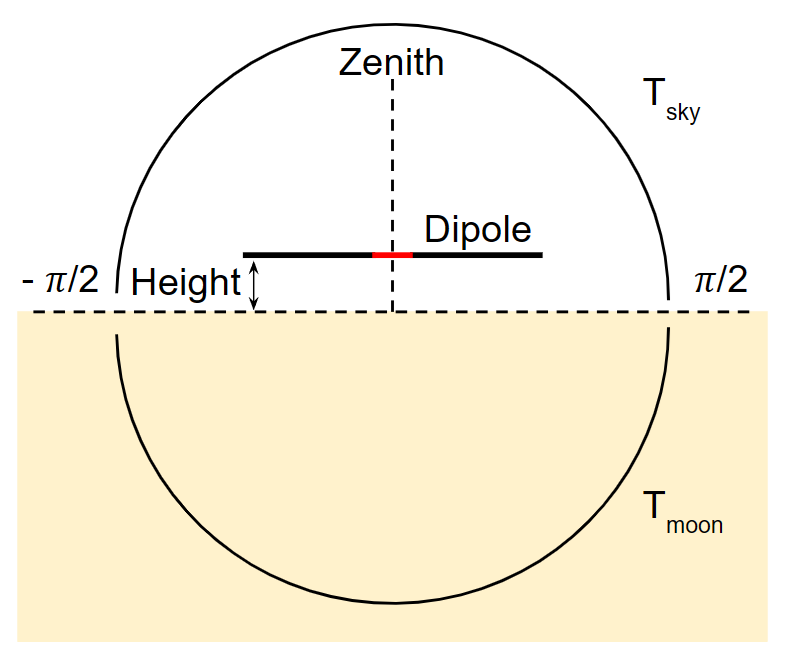}\hfill
  \includegraphics[height=4.2cm]{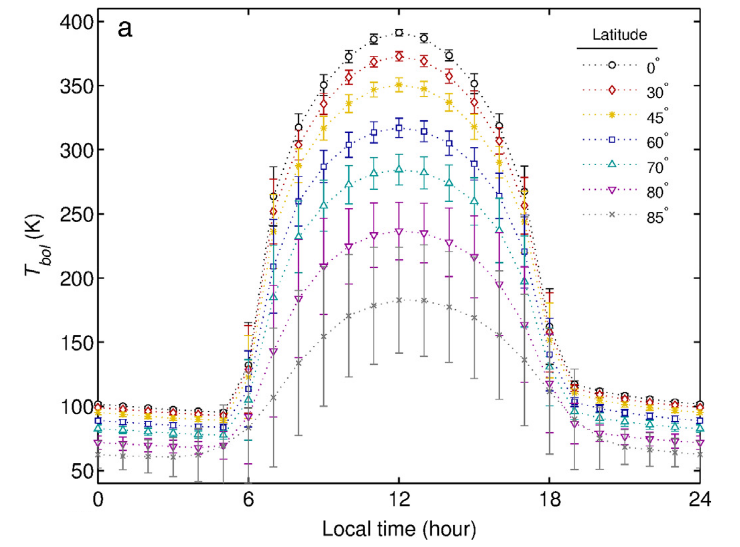}\hfill
  \includegraphics[height=4.4cm]{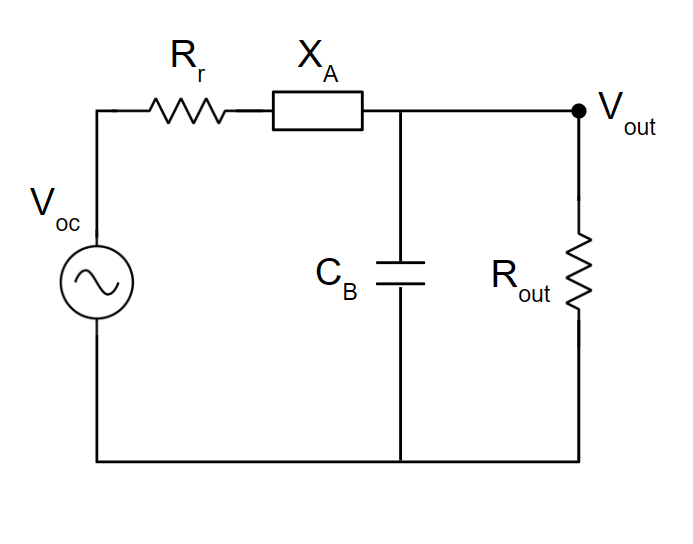}
  \caption{(\emph{Left}) Coordinate system used to calculate effective temperature seen by the antenna.
  (Middle) Temperature on the lunar surface over the course of a lunar day. Latitudes have a dramatic effect on the daytime temperature. Plot taken from \citet{Williams2017}. (\emph{Right}) Equivalent circuit model of antenna circuit. $\mathrm{R}_{\mathrm{ant}}$ and $\mathrm{X}_{\mathrm{ant}}$ are real and imaginary antenna impedance. A stray capacitance $\CB$ arises between the antenna and antenna housing, body, and cables. An output resistance of $\Rout$ = 1 M$\Omega$ was selected to maximize output voltage while minimizing Johnson noise.
  \label{fig:circuit}}
\end{figure}

Both the impedance and the beam affect how the antenna couples to the sky. 
A dipole deployed to the moon will observe both the sky and lunar ground. The antenna brightness temperature ($\TA$), the effective temperature that the dipole sees, can be calculated by integrating over the sky- and ground-hemispheres as depicted in the cartoon in the \emph{left} panel of Fig.~\ref{fig:circuit}.
The antenna brightness temperature is calculated as,
\begin{equation}
    \TA = \Tsky \fsky + \Tmoon (1-\fsky).
\end{equation}
The middle plot of Fig.~\ref{fig:circuit} shows the temperature near the surface of the moon over the course of a lunar day for various latitudes \citep{Williams2017}. While daytime temperatures vary greatly with the latitude, lunar night temperatures are much more consistent. We assume lunar temperature $\Tmoon$ of 100 Kelvin during lunar night. 
For the sky temperature $\Tsky$, we adopt the \citet{cane1979spectra} frequency-dependent model. For simplicity, we assume a spatially uniform sky temperature.

\begin{figure}
  \centering
  \includegraphics[height=5.5cm]{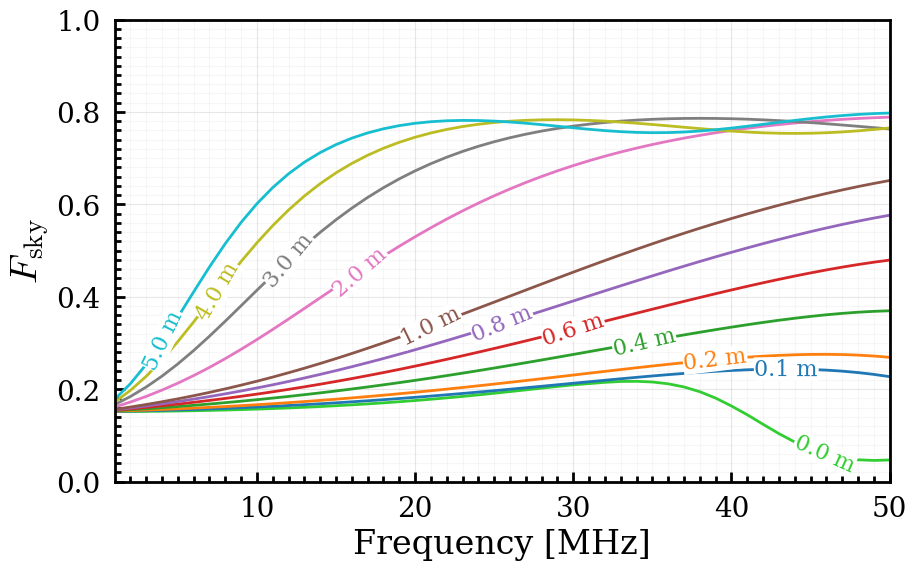} 
  \hfill
  \includegraphics[height=5.5cm]{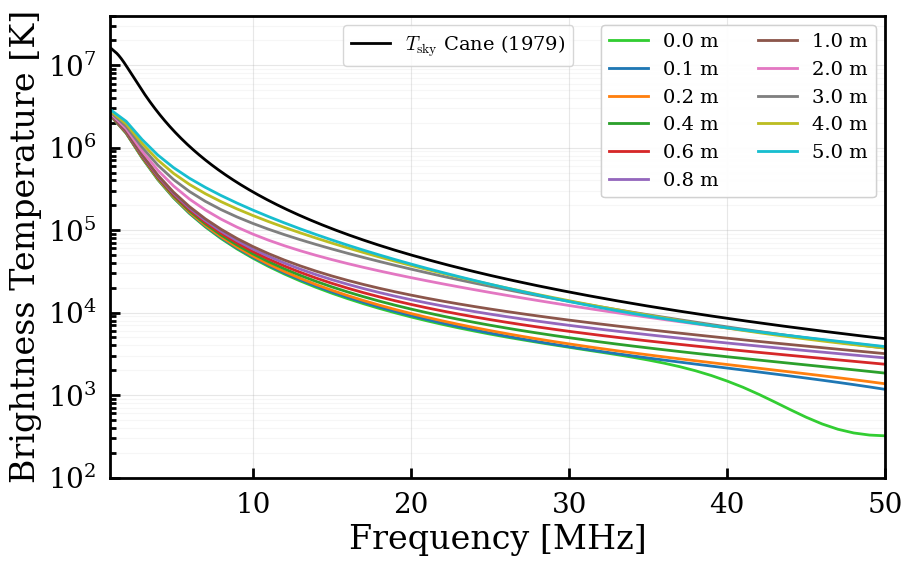}
  \caption{(\emph{Left}) Fraction of the radiated beam that sees the sky as determined by EM simulations. (\emph{Right}) Antenna brightness temperature of the dipole (colored lines) given a synchrotron-dominated sky as modeled by \citet{cane1979spectra} (black line). 
  \label{fig:Fg}}
\end{figure}

The frequency-dependent sky fraction for our dipole is shown in the \emph{left} plot of Fig.~\ref{fig:Fg}. As the dipole height is increased, the coupling of the beam to the ground diminishes and larger fraction of the gain of the antenna is aimed at the sky. The corresponding antenna brightness temperature for a synchrotron-dominated sky is shown in the \emph{right} plot of Fig.~\ref{fig:Fg}. We find that as the dipole is raised above the ground, its effective temperature tends towards that of the sky, as expected. Once again, the unique case of 0~meters bucks the trend displayed by the other heights with large deviations above 35~MHz.

To determine the anticipated sensitivity of a dipole deployed to the far-side of the moon, we envision the antennas are mounted to some structure and fed into a receiver terminal. 
We assume an equivalent circuit as shown in the \emph{right} panel of Fig.~\ref{fig:circuit}. The antenna have complex impedance $\Zant$ where the radiation resistance ($\R$) and reactance ($\X$) are the real and imaginary components, respectively. 
Following the formalism set by \citet{7776014}, the open circuit rms voltage ($\Voc$) at the antenna terminal when the antenna is immersed in a radiation field with a brightness temperature $\TA$ is given by:
\begin{equation}
    \Voc = \sqrt{4k_{B}\TA \R\Delta f},
\end{equation}
where $k_B$ is a Boltzmann constant and $\Delta f$ is the bandwidth. Using the simple voltage divider picture depicted in the equivalent circuit, we can determine the output voltage feeding into the receiver terminal:
\begin{equation}
    \Vout = \Voc \frac{|\Zrec|}{|\Zrec+\Zant|}, 
    \hspace{1cm}\mathrm{where}\hspace{1cm}
    \Zrec = \frac{1}{\frac{1}{\Rout}+i\omega \CB}
\end{equation}

The base capacitance ($\CB$) in the circuit arises from input impedance of JFET amplifier and stray capacitance between the monopole and surrounding structures such as antenna housings. 
We assumed 35~pF for this capacitance taken from the STEREO/WAVES experiment \citep{STEREO2008} that deployed stacer antennas coupled to low noise JFET amplifiers. For the parallel real impedance $\Rout$, it is straightforward to show that for resistances above 10 k$\Omega$ the sensed voltage at the input of the JFET is largely unchanged. For example, the STEREO/WAVES experiment used a 150 M$\Omega$ for this resistor \citep{STEREO2008}. For this study, we assume $\Rout$ to be 1~M$\Omega$.

\begin{figure}
  \centering
  \includegraphics[height=5.5cm]{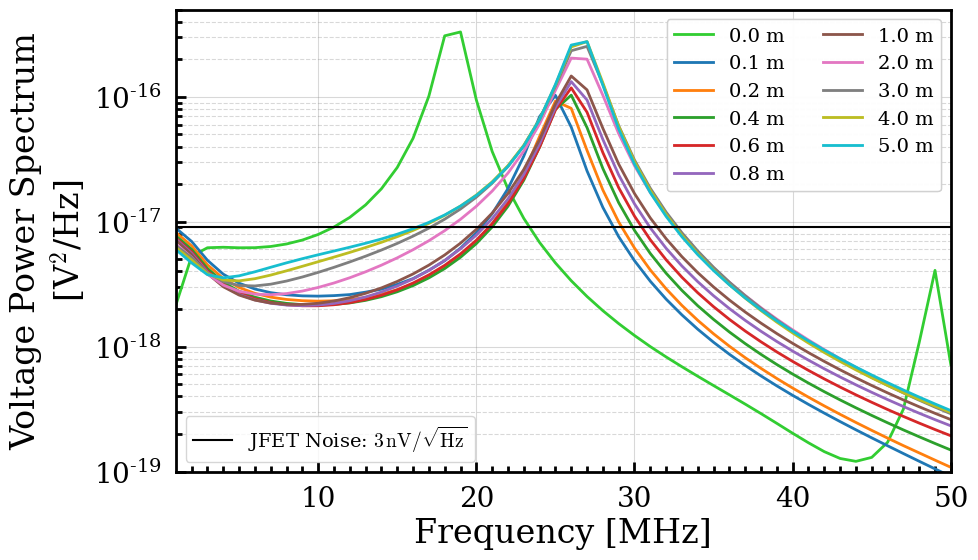}
  \caption{Voltage power at receiver terminals given the effective brightness temperature of the antennas and a synchrotron-dominant sky. Noise values are shown for reference.
  \label{fig:Vout}}
\end{figure}

The expected voltage power spectrum as the antenna observes a synchrotron-dominated sky from the surface of the moon is plotted in Fig.~\ref{fig:Vout}. 
The black horizontal line indicates an assumed  3~nV$/\sqrt{\mathrm{Hz}}$ noise floor of a JFET amplifier.
Simply put, to ensure measurements are not limited by amplifier noise, the voltage power from sky temperature (colored curves) must exceed the noise floor of the chosen amplifier (black line). 

\begin{figure}
  \centering
  \includegraphics[height=5.5cm]{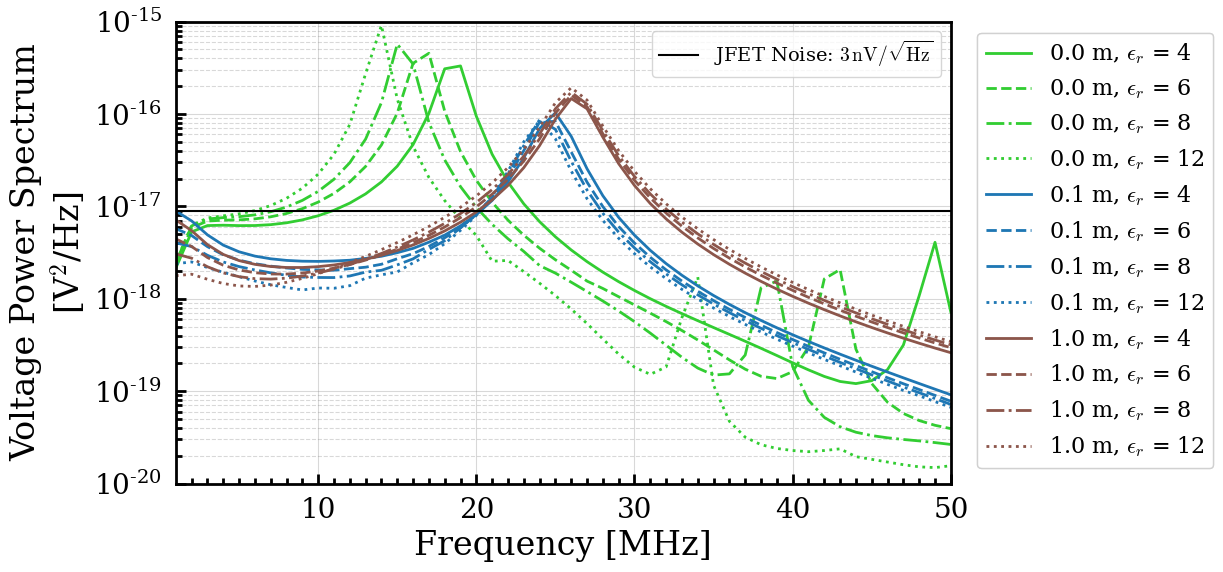}
  \caption{Voltage power at receiver terminals given the effective brightness temperature of the antennas and a synchrotron-dominant sky at 0~meter, 0.1~meter, and 1~meter high above the dielectric half-space with varying dielectric constants.
  \label{fig:Vout_dielectric}}
\end{figure}

The elevated sensitivity of the 0-meter case can be explained as follows: The antenna resonates at lower frequency due to coupling with the ground, as shown in Fig.~\ref{fig:ZA}. Because the temperature of the sky increases rapidly at lower frequencies (see Fig.~\ref{fig:Fg}), the large real component of the impedance  results in the sky noise dominating over the amplifier noise. This would then suggest that placing the antenna directly on the ground, reducing engineering challenges and costs considerable, has merit.
However, as is seen in Fig.~\ref{fig:Vout_dielectric}, when the dipole is on the ground, the voltage power spectrum is highly sensitive to the precise dielectric properties. Changes in the regolith dielectric constant will shift the frequency response of the antenna. 
Therefore uncertainties of the lunar regolith will propagate sizable deviations into the voltage power spectrum. Unfortunately, without dedicated measurements from the specific landing site, these uncertainties persist.
This soil-dependence is largely eliminated even for small dipole heights, a conclusion that argues against placing the dipole directly on the ground.
Placing the antenna higher off the ground will increase the sky signal as well as broadens the frequency range in which the sky noise surpasses the amplifier noise, therefore improving the over-all instrument sensitivity. 
However, the complexity of the skyward beam increases as the height of the antenna increases, as shown in Figure~\ref{fig:TEPower}. The number of lobes in the beam of an infinitesimally small dipole is given by: 
\begin{equation}
    \mathrm{N_{lobes}} \sim \frac{2h}{\lambda}+1,
\end{equation}
where $h$ is the distance between dipole and the conductor and $\lambda$ is the wavelength \citep{balanis2015antenna}. 
Therefore there is a trade-off between sensitivity, beam complexity, systematic uncertainty, and engineering challenges that need to be balanced for the future lunar radio astrophysics experiments. 

\section{Conclusion}
We simulated a 6-meter tip-to-tip dipole antenna at various heights above a dielectric half-space with varying dielectric constants $\epsilon_r$. Our choice in dielectric constants were based off the analysis conducted by \citet{Grimm2018} of the near-side regolith. Here we restricted ourselves to the simple scenario of a dielectric half-space with a single dielectric constant. Future work will expand the analysis to include a layered impedance, mimicking the situation of a potential dipole antenna deployed to the lunar far-side more closely. We adopt the \citet{cane1979spectra} model for a synchrotron-dominated sky, assuming spatial uniformity for simplicity. General agreement of the simulated beams with analytical models presented by \citet{Jiao} engender confidence; further analysis was based off simulations. We calculated how the simulated antenna impedance and beam for a representative lunar surface instrument exploiting the radio-quiet environment of the lunar far-side would observe the radio sky. 

For the unique case of the dipole physically touching the dielectric half-space, the voltage power spectrum and the sky-noise dominated regime have a strong dependence on the dielectric constant. 
For the simplified case presented here, a single nuisance parameter might self-calibrate this uncertainty as a part of a global data fit. However, for a realistically complex case with potentially inhomogeneous and layered regolith, this uncertainty is likely to be a limiting factor without fully characterized dielectric properties of the landing site. We find that even small elevations of the dipole over the ground alleviates these uncertainties. 

For the antenna raised off the ground, the coupling to the sky increases as the dipole is raised higher, with the added benefit of broadening the frequency range over which the  sensitivity is sufficient to be sky-noise dominated. We also saw that the antenna impedance becomes less sensitive to the properties of the regolith. 
However, the benefits of improved sky-sensitivity with height are countered by increased beam complexities and engineering challenges of realizing a taller structure on the moon. 

A balance must be struck among antenna sensitivity, signal to noise ratio, beam chromaticity, systematic uncertainties, and engineering constraints, including cost and mass.

\section{Acknowledgments}
Support for LuSEE-Night efforts at UC Berkeley is provided by NASA contract 80MSFC23CA015.  SDB acknowledges support of the Royal Society Wolfson Visiting Fellowship program.
KMR, A Suzuki and A Slosar acknowledge support for LuSEE-Night Science by Higher Energy Physics, Office of Science, US Depamertment of Energy.
\bibliographystyle{aasjournalv7}
\bibliography{refs}

\end{document}